\documentclass[a4paper,fleqn,usenatbib]{mnras}

\usepackage{newtxtext,newtxmath}

\usepackage[T1]{fontenc}
\usepackage{ae,aecompl}

\usepackage{graphicx}	
\usepackage{amsmath}	
\usepackage{amssymb}	

\usepackage{adjustbox} 
\usepackage{lipsum}

\newcommand{\Me}{M$_\oplus$}
\newcommand{\Mjup}{M$_\mathrm{Jup}$}
\newcommand{\Msun}{M$_{\sun}$}

\newcommand{\Mstar}{\mathrm{M}_\star}
\newcommand{\ap}{a_\mathrm{p}}
\newcommand{\ep}{e_\mathrm{p}}
\newcommand{\Mp}{M_\mathrm{p}}

\newcommand{\Md}{M_\mathrm{d}}
\newcommand{\rmin}{r_{\min}}
\newcommand{\rmax}{r_{\max}}

\newcommand{\dgap}{\delta_\mathrm{g}}

\newcommand{\wgap}{w_\mathrm{g}}
\newcommand{\rgap}{r_\mathrm{g}}
\newcommand{\ed}{e_\mathrm{d}}

\newcommand{\uJybeam}{$\mu$Jy~beam$^{-1}$}
\newcommand{\kms}{km~s$^{-1}$}

\title[ALMA observations of HD~92945]{A gap in HD~92945's broad
  planetesimal disc revealed by ALMA}
\author[S. Marino et al.]{ S. Marino$^{1,2}$\thanks{E-mail:
    sebastian.marino.estay@gmail.com}, B. Yelverton$^{1}$, M. Booth$^{3}$, V. Faramaz$^{4}$, G. M. Kennedy$^{5,6}$, L. Matr{\`a}$^{7}$\newauthor{and M. C. Wyatt$^{1}$} \\
  $^{1}$Institute of Astronomy, University of Cambridge, Madingley Road, Cambridge CB3 0HA, UK\\
  $^{2}$Max Planck Institute for Astronomy, K\"onigstuhl 17, 69117 Heidelberg, Germany\\
  $^{3}$Astrophysikalisches Institut und Universit\"atssternwarte, Friedrich-Schiller-Universit\"at Jena, Schillerg\"a\ss{}chen 2-3, 07745 Jena, Germany\\
  $^{4}$Jet Propulsion Laboratory, California Institute of Technology, 4800 Oak Grove drive, Pasadena CA 91109, USA.\\
  $^{5}$Department of Physics, University of Warwick, Gibbet Hill Road, Coventry, CV4 7AL, UK\\
  $^{6}$Centre for Exoplanets and Habitability, University of Warwick, Gibbet Hill Road, Coventry, CV4 7AL, UK\\
  $^{7}$Harvard-Smithsonian Center for Astrophysics, 60 Garden Street, Cambridge, MA 02138, USA\\
}

\date{Accepted XXX. Received YYY; in original form ZZZ}

\pubyear{2018}

\begin{document}
\label{firstpage}
\pagerange{\pageref{firstpage}--\pageref{lastpage}}
\maketitle

\begin{abstract}
  In the last few years multiwavelength observations have revealed the
  ubiquity of gaps/rings in circumstellar discs. Here we report the
  first ALMA observations of HD~92945 at 0.86~mm, that reveal a gap at
  about $73\pm3$~au within a broad disc of planetesimals that extends
  from 50 to 140~au. We find that the gap is $20^{+10}_{-8}$~au
  wide. If cleared by a planet in situ, this planet must be less
  massive than 0.6~\Mjup, or even lower if the gap was cleared by a
  planet that formed early in the protoplanetary disc and prevented
  planetesimal formation at that radius. By comparing opposite sides
  of the disc we also find that the disc could be
  asymmetric. Motivated by the asymmetry and the fact that planets
  might be more frequent closer to the star in exoplanetary systems,
  we show that the gap and asymmetry could be produced by two planets
  interior to the disc through secular resonances. These planets
  excite the eccentricity of bodies at specific disc locations,
  opening radial gaps in the planetesimal distribution. New
  observations are necessary to confirm if the disc is truly
  asymmetric, thus favouring the secular resonance model, or if the
  apparent asymmetry is due to a background galaxy, favouring the
  in-situ planet scenario. Finally, we also report the non-detection
  of CO and HCN gas confirming that no primordial gas is present. The
  CO and HCN non-detections are consistent with the destruction of
  volatile-rich Solar System-like comets.
\end{abstract}





\section{Introduction}
\label{sec:intro}

In the last decade, observational campaigns have discovered thousands
of exoplanets with orbital radii smaller than 5 au; however, the
population of planets at larger radii remains mostly unexplored,
especially for low mass planets \citep{Bowler2018}. Those planets,
nevertheless, can shape the distribution of planetesimals leaving
hints of their presence. Our Solar System is a good example, with the
Kuiper belt's inner edge shaped by Neptune, and the gap between the
asteroid and Kuiper belt filled with planets.

In this manner, observations of dusty debris discs can reveal the
presence of planets that shape the distribution of planetesimals
\citep[potentially stirring their orbits, e.g.][]{Mustill2009} and
dust \citep[see reviews by][]{Wyatt2008, Hughes2018}. Well known
examples of systems with debris discs and confirmed planets are
$\beta$~Pic \citep{Lagrange2009}, HR~8799 \citep{Marois2008,
  Marois2010, Booth2016, Read2018, Wilner2018} and HD~95086
\citep{Rameau2013, Su2013, Su2015, Rameau2016}. Other systems display
eccentric debris discs that hint at the presence of yet undetected
planets \citep[e.g. Fomalhaut and HD~202628,][Faramaz et al. in
  prep]{Kalas2005, Krist2012, Schneider2016}; or both inner and outer
belts \citep[e.g.][]{Backman2009, Morales2009, Chen2009,
  Ballering2014, Kennedy2014}, which have been used as an argument for
the presence of planets in between these two components
\citep{Shannon2016, Lazzoni2018, Matthews2018}. Additionally, a few
debris discs show evidence of single or multiple gaps in between broad
outer belts which suggests the presence of planets formed within these
icy planetesimals belts. The most remarkable of these systems is
HD~107146, with a face on debris disc around a solar analogue, 100~au
wide, and with a $\sim50$~au wide gap at 80~au revealed by ALMA
\citep{Ricci2015, Marino2018hd107}.


\begin{table*}
  \centering
  \caption{Summary of band 7 (12m and ACA) observations. The image rms
    and beam size reported correspond to natural weighting.}
  \label{tab:obs}
  \begin{adjustbox}{max width=1.0\textwidth}
    \begin{tabular}{lrccccccc} 
  \hline
  \hline
  Observation & Dates              & t$_\mathrm{sci}$  & Image rms  & beam size (PA) & Min and max baselines [m]  \\
                &                  &      [hours]       & [$\mu$Jy]  &                &  (5th and 95th percentiles)   \\
  \hline
  Band 7 - 12m & 13, 17, 18 Dec 2016    & 2.2  & 16 & $0\farcs55\times0\farcs48$ ($73\degr$) & 46 and 360  \\
  Band 7 - ACA & 24-27 Oct 2016   & 3.6 &  175 &  $5\farcs1\times3\farcs0$ ($-84\degr$) & 9 and 44 \\
               & 23 Mar 2017  & &&&&&& \\
  Band 7 - 12m+ACA & - & - & 16 & $0\farcs54\times0\farcs49$ ($69\degr$) & - \\
    \hline
  \end{tabular}
  
  \end{adjustbox}
\end{table*}

Two other systems (HD~131835 and HD~92945) also show tentative
evidence of gaps or concentric rings possibly produced by planet disc
interactions \citep{Feldt2017, Golimowski2011,
  Schneider2014}. However, these features have only been detected in
scattered light observations that trace the distribution of
$\mu$m-sized dust grains, which are heavily affected by radiation
forces, thus not directly tracing the distribution of the parent
planetesimals. Moreover, HD~131835 is known to have large amounts of
gas \citep{Moor2015gas, Kral2018CI} which could affect the
distribution of $\mu$m-sized grains traced in scattered light
observations, and cause instabilities that can result in multiple ring
structures \citep{Klahr2005, Lyra2013, Richert2017}.

In this paper we focus on HD~92945, a 100-300~Myr old K0 star
\citep{Song2004, Plavchan2009} at 21.5~pc \citep{Gaiadr2}, and present
the first ALMA observations of this system to search for evidence of
single or multiple planets that could be shaping the distribution of
planetesimals. ALMA observations can resolve emission arising from
0.1-10~mm-sized grains, for which radiation forces are negligible, and
thus trace the location of the parent planetesimals.

This paper is organised as follows: In \S\ref{sec:obs} we present the
first ALMA observations of HD~92945, revealing a broad planetesimal
disc with a gap near 73~au. Then, in \S\ref{sec:parmodel} we model the
observations to characterise the distribution of planetesimals,
especially the width and depth of the gap, and test if the disc could
be asymmetric. Given the derived disc structure, in \S\ref{sec:SR} we
propose that a pair of inner planets could produced the observed gap
through secular resonances. In \S\ref{sec:dis} we discuss potential
origins for the observed gap, and which scenarios we can already rule
out. Finally, in \S\ref{sec:conclusions} we summarise the main
findings and conclusions of this work.




\section{Observations}
\label{sec:obs}

 \begin{figure*}
  \centering
  \includegraphics[trim=0.0cm 0.0cm 0.0cm 0.0cm, clip=true, width=0.45\textwidth]{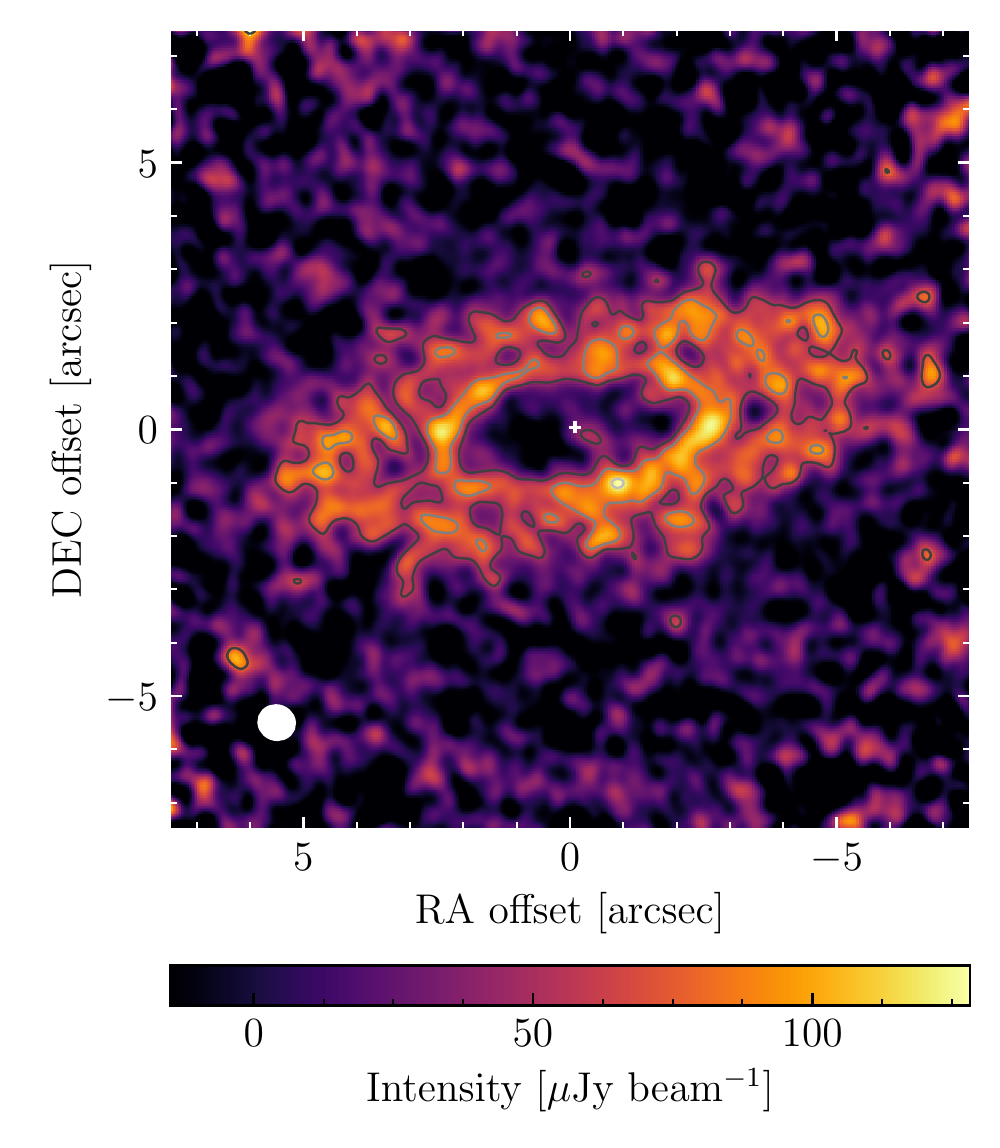}
   \includegraphics[trim=0.0cm 0.0cm 0.0cm 0.0cm, clip=true, width=0.45\textwidth]{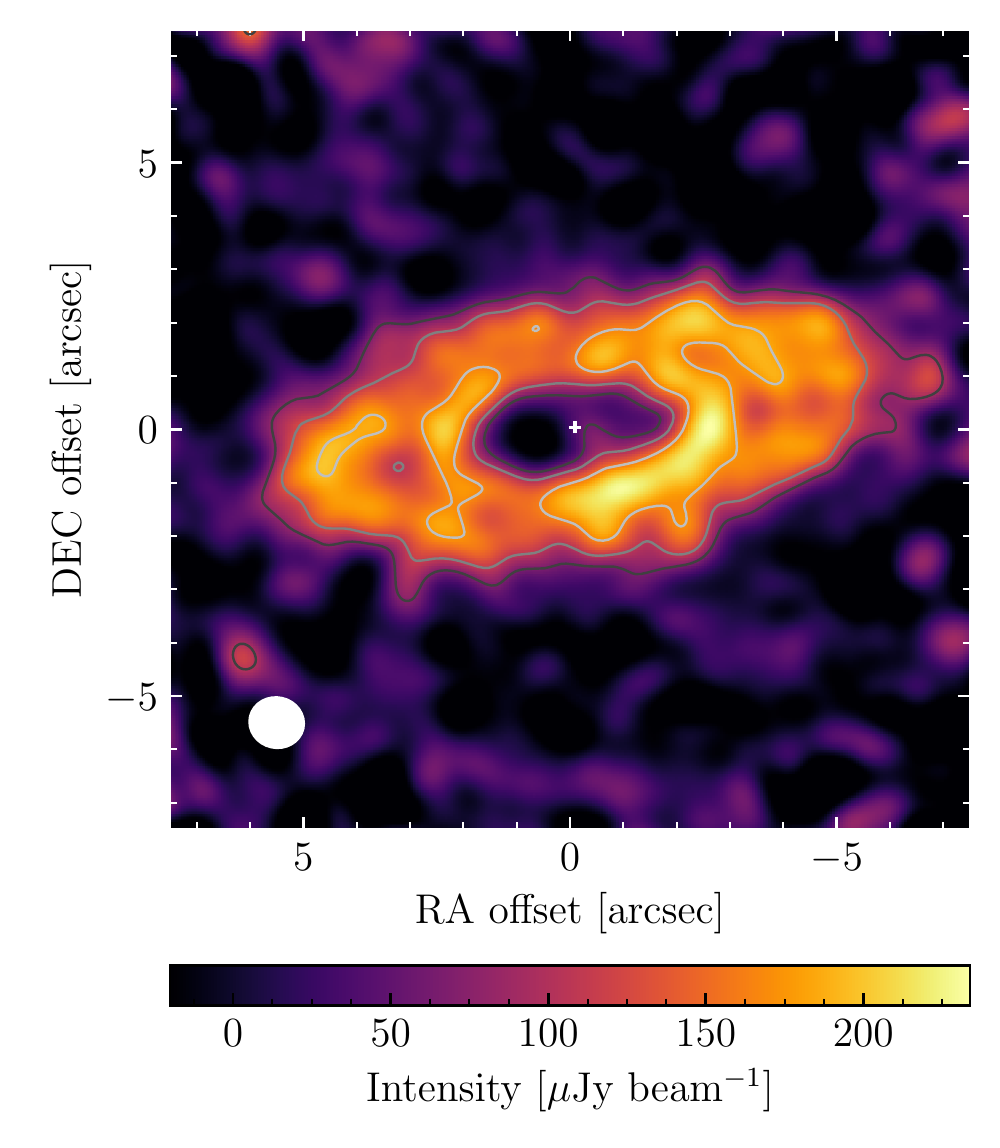}
 \caption{Continuum Clean images at 0.86mm (12m+ACA) of HD~92945
   obtained using natural weights (left panel) and an additional uv
   tapering of $0\farcs7$ (right panel). The images are also corrected
   by the primary beam, hence the noise increases towards the
   edges. The contours represent 3, 5 and 8 times the image rms (16
   and 20~\uJybeam\ at the center of the images without and with
   uvtapering respectively). The stellar position is marked with a
   white cross near the center of the image (based on Gaia DR2) and
   the beams are represented by white ellipses in the bottom left
   corners ($0\farcs54\times0\farcs49$ and $0\farcs87\times0\farcs79$
   without and with uvtapering respectively). }
 \label{fig:alma}
\end{figure*}

 \begin{figure}
  \centering \includegraphics[trim=0.0cm 0.0cm 0.0cm 0.0cm, clip=true,
    width=1.0\columnwidth]{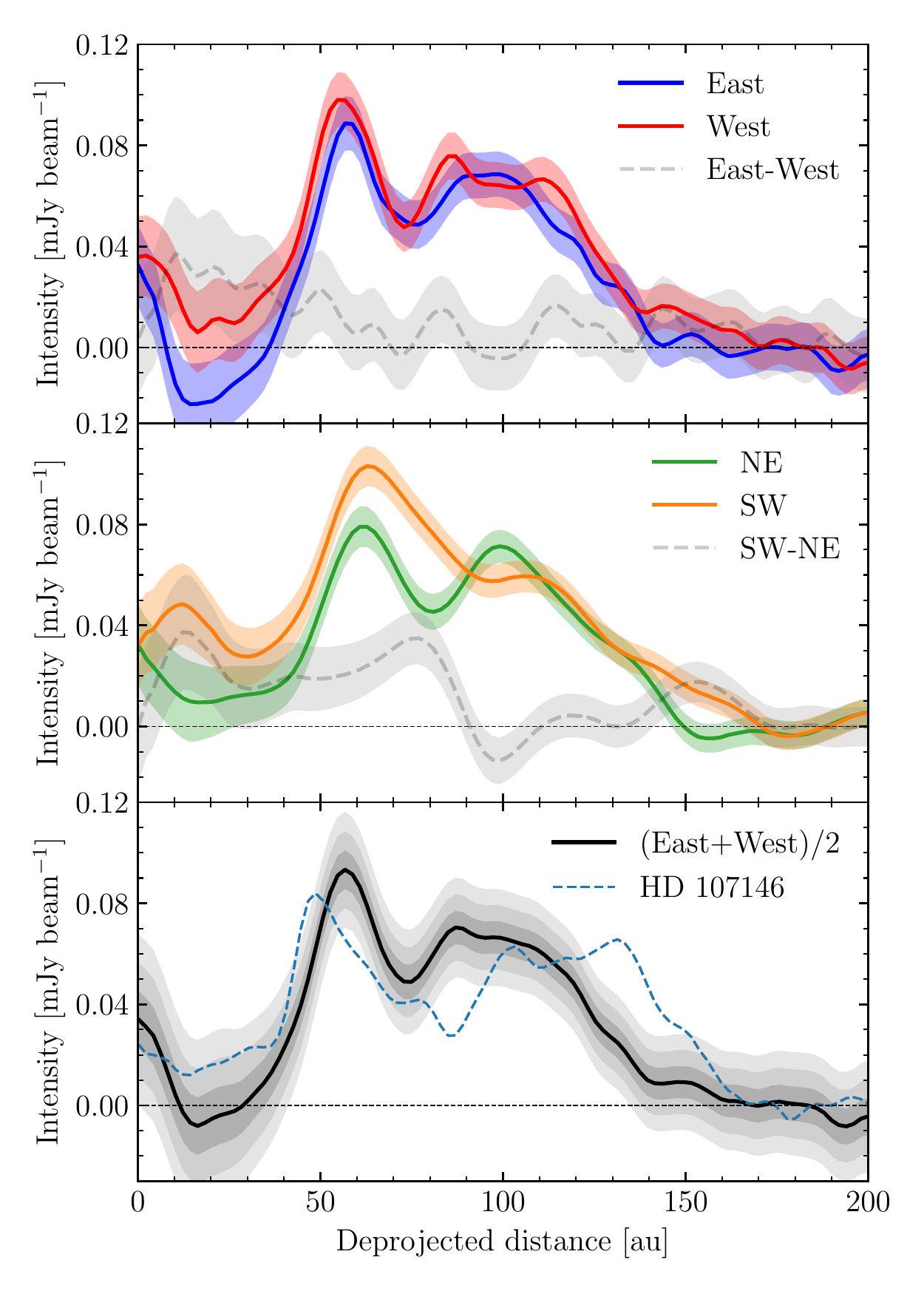}
 \caption{Deprojected surface brightness profile computed by
   azimuthally averaging the emission (Clean image with natural
   weights) within $60\degr$ wide wedges. \textbf{\textit{Top:}}
   Comparison between the East (blue) and West (red) side of the disc
   along the disc major axis, i.e. at a mean PA of $100\degr$ and
   $280\degr$, respectively. \textbf{\textit{Middle:}} Comparison
   between opposite wedges with a mean PA of $45\degr$ (NE, green) and
   $225\degr$ (SW, orange). The shaded regions in the top and middle
   panels correspond to $1\sigma$
   uncertainties. \textbf{\textit{Bottom:}} Average of the East and
   West sides. The shaded regions correspond to 1, 2 and $3\sigma$
   uncertainties. The grey dashed lines represent the difference
   between the East and West wedges (top) and SW and NE wedges
   (middle). Note that the shaded regions are representative of the
   uncertainty over a resolution element, i.e. 12~au for the top and
   bottom panels and 20~au for the middle panel approximately. The
   blue dashed line in the bottom panel represents the averaged radial
   intensity profile of HD~107146 in band 7, with a resolution of
   13~au and avoiding the South East side of the disc where a clump is
   detected \citep{Marino2018hd107}.}
 \label{fig:radial_profile}
\end{figure}

We observed HD~92945 in band 7 (0.86~mm) as part of the cycle 4
project 2016.1.00104.S (PI: S. Marino). We requested both 12m array
and ACA observations to obtain high angular resolution and to recover
large scale structure that was previously observed by HST
\citep{Golimowski2011, Schneider2014}. The 12m array observations were
carried out between 13 and 18 December 2016, for a total of 2.2~h on
source. The baselines ranged from 46 to 360~m (5th and 95th
percentiles), with PWV values between 0.6 and 1.0~mm and a total of
42-47 antennas were used. ACA observations were taken between 24
October 2016 and 23 March 2017, with a total time on source of 3.6~h
and PWV values ranging from 0.3 to 1.4~mm. The baselines ranged from 9
to 44~m and 8-11 antennas were used. These details are summarised in
Table \ref{tab:obs}. This range of baselines allows us to recover
emission on angular scales ranging from $0\farcs5$ to $20\arcsec$ or
10 to 400~au.

The correlator was set up with four spectral windows. Two were used to
study the continuum emission only, centred at 343.1 and 357.0~GHz
with 2~GHz bandwidth and a channel width of 15.625 MHz. The other two
were center-ed at 345.0 and 355.1 to look for CO and HCN gas emission,
with 2~GHz bandwidth and channel widths of 0.977~MHz (effective
spectral resolution of 1.13 MHz) and 0.488~MHz (effective spectral
resolution of 0.977 MHz), respectively. Calibrations were applied
using CASA 4.7 \citep{casa} and the calibration routines provided by
ALMA.

\subsection{Dust continuum}
\label{sec:continuum}

Continuum maps at 0.86~mm are obtained using the \textsc{tclean} task
in CASA 5.1 \citep{casa}. In Figure \ref{fig:alma} we present the
Clean image using natural weights (left panel) and smoothed with a
Gaussian tapering of $0\farcs7$ (right panel) to obtain a higher
signal-to-noise per beam (but at lower resolution). These images
reveal a well resolved broad disc with a bright inner rim at
$2\farcs5$ (54~au) from the star and a fainter outer component
extending out to $\sim7\arcsec$ (150~au) with a sharp outer edge,
similar to the morphology seen by HST \citep{Golimowski2011,
  Schneider2014}. The new ALMA image indicates that large grains and
thus planetesimals are present at a wide range of radii in a
$\sim100$~au wide disc, and not only at the \textit{bright inner ring}
around 50~au as it was modelled by \cite{Golimowski2011}. This is not
always the case, with many systems displaying narrow birth rings well
traced at mm wavelengths, with extended halos of small grains blown
out by radiation pressure \citep[e.g. as seen for HD~181327, Fomalhaut
  and HD~202628 in scattered light,][Faramaz et al. in
  prep]{Schneider2014, Stark2014, Marino2016, Boley2012}. The
similarity between scattered light and mm thermal emission images
could be due to HD~92945 being a K0 star, thus $\mu$m-sized grains are
not as heavily affected by radiation pressure as around more luminous
stars.

We measure a total disc flux of $9.6\pm1.0$~mJy (including absolute
flux uncertainties) by integrating the emission inside an elliptic
mask oriented the same as the disc in the sky (see
\S\ref{sec:parmodel}) and with a semi-major axis of $7\arcsec$
(150~au). This total flux is consistent with the $8.6\pm1.1$~mJy flux
measured with JCMT/SCUBA-2 as part of the SONS survey
\citep{Panic2013, Holland2017}. Within the inner edge of the disc, a
$3\sigma$ peak ($\sim50~\mu$Jy) appears slightly offset by
$0\farcs25$, but is still consistent with being emission arising from
the star given the astrometric accuracy of $0\farcs16$\footnote{ALMA
  Technical Handbook, Astrometric observations.} and the expected
photospheric emission ($30~\mu$Jy based on photometry at wavelengths
shorter than 10~$\mu$m and assuming a Rayleigh-Jeans spectral
index). We also find that the disc inner edge is brighter towards the
South-West than elsewhere, indicating that the disc might be
asymmetric, as suggested by HST images \citep{Golimowski2011,
  Schneider2014}, although those scattered light observations showed
the East side being brighter than its counterpart.

Despite the low S/N per beam in these images, it is still possible to
see along the disc major axis that the surface brightness has a local
minimum at about $3\farcs4$ (73~au) from the center, which suggests
the presence of a gap. This finding seems to confirm the previous
suggestion of a gap at the same location based on scattered light
images presented by \cite[see their Figure 7 for a dust density radial
  profile]{Golimowski2011}. This feature is better seen in a disc
brightness radial profile, obtained by azimuthally averaging the
deprojected emission (using the disc inclination and PA found in
\S\ref{sec:parmodel}) within $60\degr$ wide wedges as shown in Figure
\ref{fig:radial_profile}. The radial profile strongly peaks at around
55~au, with no significant emission within 40~au, and with a local
minimum near 73~au that is seen both at the Eastern and Western sides
of the disc along the disc major axis. These wedges are consistent
with each other; however, when comparing the radial profiles obtained
along a PA of $45\degr$ and $225\degr$ (NE and SW lines in Figure
\ref{fig:radial_profile}) we find $3.4\sigma$ differences. The
South-West side is brighter than its counterpart and shows no clear
evidence of a gap (orange line). These findings suggest that the disc
is asymmetric, providing key information to understand the origin of
this gap (see \S \ref{sec:SR} and \ref{sec:dis}).

It is also noticeable some similarities in the radial structure of the
debris discs around HD~92945 and HD~107146 (see bottom panel of Figure
\ref{fig:radial_profile}). Both are among the widest discs that have
been imaged in the sub-millimetre \citep{Matra2018mmlaw}.  HD~107146's
outer belt extends from roughly 40-150~au, with a wide gap centred at
around 75~au and bright and sharp disc inner and outer edges. In
\S\ref{sec:parmodel} we conclude a similar radial structure for
HD~92945, although with an overall narrower disc and gap. The stellar
masses and ages of these systems are also similar (0.9-1.0 \Msun\ and
100-300 Myr). However, the HD~92945 disc has a dust mass that is a
factor 5 lower (which also applies to the total disc mass if both have
the same planetesimal size distribution), and moreover, is possibly
asymmetric. The difference in mass could still be consistent with
similar initial masses, but more depletion over time for HD~92945 than
HD~107146 if the first is older by a factor of a few, or the largest
planetesimals are larger in the case of HD~92945. Alternatively, if
the level of stirring is higher in HD~92945 it would collisionally
evolve faster, explaining its lower mass.

\subsection{CO J=3-2 and HCN J=4-3}
\label{sec:gas}

As well as studying the dust continuum emission we look for any
secondary gas that could be present, released from volatile rich
planetesimals in the disc \citep[e.g.][]{Dent2014, Marino2016,
  Matra2017betapic, Matra2017fomalhaut}. However, after subtracting
the continuum emission and producing dirty data cubes (no clean
iterations are necessary) we find no evidence for CO (3-2) nor HCN
(4-3) line emission co-located with the dust, nor closer in as found
around $\eta$~Corvi \citep{Marino2017etacorvi}. In order to place
tight upper limits on the total CO and HCN flux, we integrate the
dirty data cubes between the disc inner and outer edge and only in
those channels where gas emission is expected according to the Doppler
shift due to Keplerian rotation, also taking the disc orientation into
account. We have used this method previously to both detect faint
lines \citep{Marino2016, Marino2017etacorvi, Matra2017fomalhaut} and
place tight upper limits \citep[e.g.][]{Matra2015}. We obtain
$3\sigma$ upper limits of 69 mJy~\kms\ for CO (3-2) and 63
mJy~\kms\ for HCN (4-3). We use these in \S\ref{sec:gasmass} to derive
gas mass upper limits.

\section{Parametric disc model}
\label{sec:parmodel}

In this section we model the dust continuum observations with a
parametric model to constrain the disc structure. We combine radiative
transfer simulations and an MCMC fitting procedure in visibility space
\citep[e.g.][]{Marino2016}. We use the same parametric model as
\cite{Marino2018hd107} that was used to fit ALMA observations of
HD~107146 (we direct the reader to that work for details). The surface
density is defined as a triple power law to constrain the disc inner
edge \textit{slope} $\gamma_1$ (power law index), the slope of the
surface density between the inner and outer edge $\gamma_2$, and the
slope beyond the disc outer edge $\gamma_3$. In addition, the model
has a gap with a Gaussian radial profile that is parametrized with a
fractional depth $\dgap$, mean radius $\rgap$ and FWHM $\wgap$.  We
also model the disc vertical aspect ratio $h$ assuming a Gaussian
vertical profile (imposing $h>0.02$ due to resolution constraints). We
leave the stellar flux at 0.86~mm ($F_\star$) as a free parameter as
it is not well constrained (e.g. it could have significant
chromospheric emission), although to compute the dust temperature we
use a stellar template spectrum corresponding to an effective
temperature of 5250~K and a radius of 0.77~$R_{\odot}$. Finally, we
use six parameters that determine the disc orientation in the sky: a
disc inclination (measured from face-on), PA, and RA and Dec offsets
(which are allowed to be different for the 12m and ACA observations).

Given the putative asymmetry found in \S\ref{sec:continuum}, we also
allow the disc to have a global eccentricity $\ed$ and argument of
pericentre $\omega$, defined as the angle in the disc plane between
the disc PA (i.e. the line of nodes) and pericentre (increasing
anti-clockwise). In this work, we use a slightly different
parametrization for the disc eccentricity since the model used in
\cite{Marino2017etacorvi} and \cite{Marino2018hd107} did not account
properly for the disc eccentricity as it assumed the disc was not
resolved radially, overestimating slightly the surface density at
apocentre. Here, we provide a correct expression for it.

First, let us assume that the disc is composed of particles with the
same eccentricity $e$ and longitude of pericentre at
$\phi=0$. Consider now a portion of the disk where semi-major axes go
from a to $a+da$ at longitudes from $\phi$ to $\phi+d\phi$. That area
is $dA=r dr d\phi$ and contains a number of particles $dN=n(a)da
d\phi/ (t_{\rm per}\dot{\phi})$, where $t_{\rm per}$ is the orbital
period, resulting in a surface density
\begin{equation}
\Sigma=\frac{dN}{dA}=\frac{n(a)}{2\upi a \sqrt{1-e^2}}. \label{eq:sigmaphi}
\end{equation}
At a given longitude $\phi$, a maps directly onto $r$ through
\begin{equation}
  a=r \left(\frac{1+e\cos(\phi)}{1-e^2}\right). \label{eq:a}
\end{equation}
This means that the surface density is constant along ellipses defined
by Equation \ref{eq:a}, and that plots of surface density versus
distance at each longitude are identical except for a scaling in
radius given by Eq. \ref{eq:a}. One way of describing this is to
define $\Sigma_r(a)=n(a)/(2\upi a)$, where $a=a(r, \phi)$ and
$\Sigma_r(a)$ is the surface density that would have resulted from the
given semi-major axis distribution for circular orbits, and otherwise
is the distribution that is scaled in radius at different longitudes
through Equation \ref{eq:a}. In our case $\Sigma_r(a)$ is parametrized
as a triple power law with a Gaussian gap as written in Equations 1
and 2 of \cite{Marino2018hd107}, but substituting $r$ by $a$ using
Equation \ref{eq:a}.

Equation \ref{eq:sigmaphi} implies that the surface density of a disc
with a global eccentricity and longitude of pericentre is independent
of $\phi$ (when keeping $a$ fixed)\footnote{We have checked if
  Equation~\ref{eq:sigmaphi} gives the right surface density
  distribution as a function of $\phi$ and $r$ by comparing it with
  numerical simulations of particles distributed in a disc on real
  eccentric orbits with semi-major axes distributed as $N(a)$. We
  found a good agreement between the two for eccentricities from 0 to
  0.9 with differences only due to numeric noise that was lower than
  1\%.}. This result is counter-intuitive because discs are expected
to be brightest at apocentre at long wavelengths since particles spend
more time near there than at any other point of their orbits
\citep{Wyatt1999, Pan2016}. However, this effect is balanced by the
increased separation between orbits (i.e.  disc radial width) at
apocentre, keeping the surface density constant. On top of this, the
disc temperature decreases with radius, making the disc slightly
brighter at pericentre. Apocentre glow should only be visible if the
emission is integrated radially or if the disc width is unresolved by
the resolution of the observations. A caveat in this derivation is
that we are assuming that all particles have the same eccentricity and
pericentre orientation, and therefore it is possible that the surface
density could be higher at apocentre under a different set of
assumptions.


\subsection{Results}

\begin{figure}
  \centering \includegraphics[trim=0.0cm 0.0cm 0.0cm 0.0cm, clip=true,
    width=1.0\columnwidth]{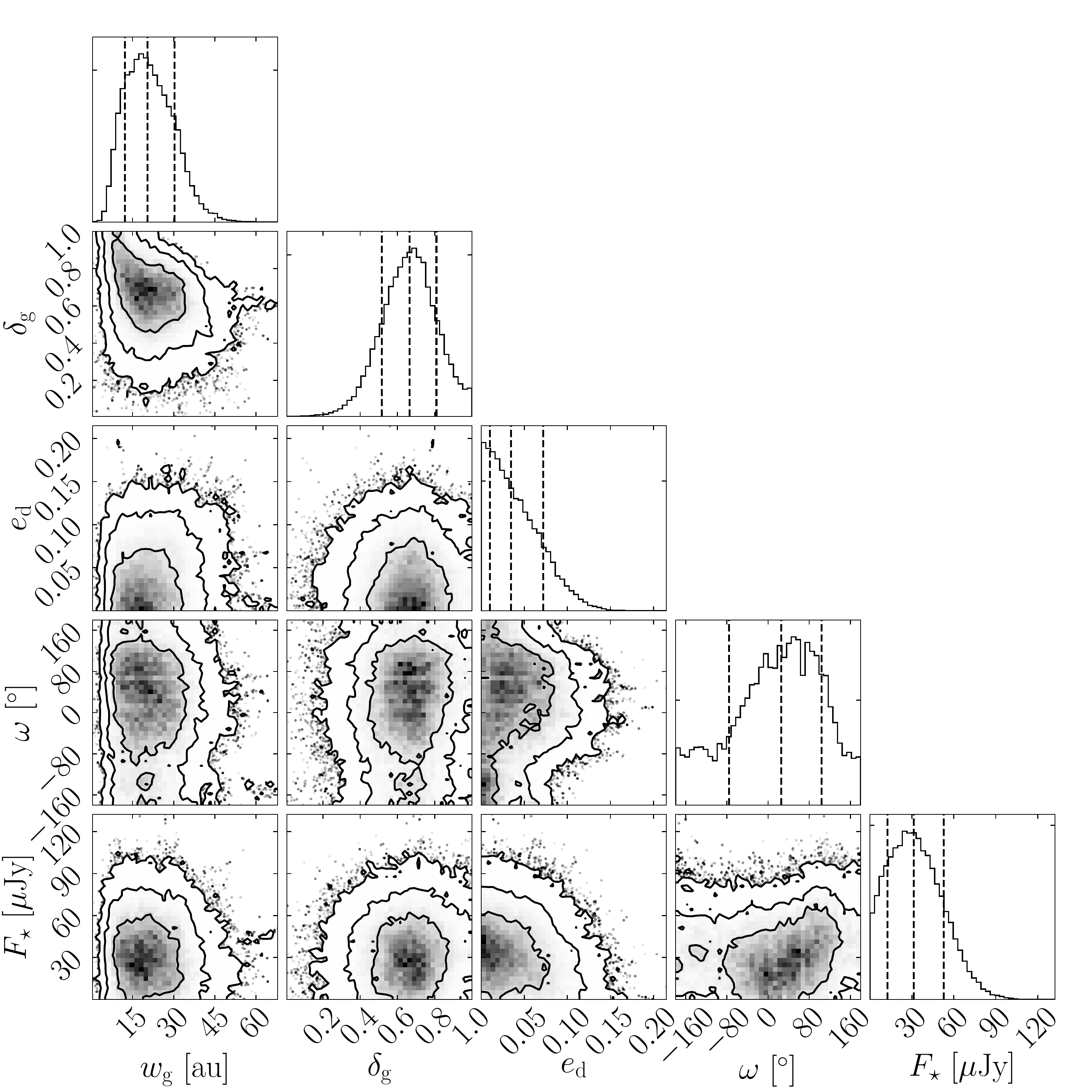}
 \caption{Posterior distributions of $\wgap$, $\dgap$, $\ed$, $\omega$
   and $F_\star$. The vertical dashed lines represent the 16th, 50th
   and 84th percentiles. Contours correspond to 68 per cent, 95 per
   cent and 99.7 per cent confidence regions. This plot was generated
   using the \textsc{PYTHON} module \textsc{CORNER} \citep{corner}.}
 \label{fig:corner}
\end{figure}

The best fit parameters are presented in Table~\ref{tab:mcmc}, and in
Figure~\ref{fig:corner} we show the posterior distribution of some of
the parameters. We find disc inner and outer radii at $52\pm3$ and
$121^{+6}_{-7}$~au, with a sharp inner (outer) edge or slope greater
(lower) than 5.7 (-6.0) with 95\% significance. The surface density
between the inner and outer disc radii is consistent with being flat
($\gamma_2=-0.52^{+0.40}_{-0.50}$), which is expected for a broad disc
of planetesimals with a surface density proportional to $r^{-1.5}$ and
a size distribution that is not yet in collisional equilibrium
\citep[i.e. the age of the system is shorter than the collisional
  lifetime of the largest planetesimals,][]{Schuppler2016,
  Marino201761vir, Geiler2017}.

\begin{table}
  \centering
  \caption{Best fit parameters of the ALMA data using our parametric
    models. The quoted values correspond to the median, with
    uncertainties based on the 16th and 84th percentiles of the
    marginalised distributions or upper limits based on 95th
    percentile.}
  \label{tab:mcmc}
  \begin{tabular}{lrl} 
    \hline
    \hline
    Parameter & best fit value & description\\
    \hline
    $\Md$ [$M_{\earth}$]& $0.047\pm0.003$ & total dust mass \\
    $\rmin$ [au] & $52\pm3$ & disc inner radius \\
    $\rmax$ [au] & $121^{+6}_{-7}$ & disc outer radius \\
    $\gamma_1$   & $>5.7$ & inner edge's slope\\
    $\gamma_2$   & $-0.52^{+0.40}_{-0.50}$ & disc slope\\
    $\gamma_3$   & $<-6.0$ & outer edge's slope \\
    $h$          & $0.049\pm0.017$ & scale height\\
    $F_\star$ [$\mu$Jy] & $30^{+21}_{-18}$ & stellar flux at 0.86 mm \\
    \hline
    $\rgap$ [au] & $73.4^{+2.7}_{-2.4}$ & radius of the gap \\
    $\wgap$ [au] & $20^{+10}_{-8}$& FWHM of the gap\\
    $\dgap$      & $0.66\pm0.15$ & fractional depth of the gap\\
    \hline
    PA [$\degr$] & $100.0\pm0.9$& disc position angle \\
    $i$ [$\degr$]& $65.4\pm0.9$ & disc inclination from face-on\\
    \hline
    $\ed$ &  <0.097 &  disc global eccentricity\\
    $\omega$ [$\degr$] & $25^{+78}_{-90}$ & argument of pericentre \\
    \hline
  \end{tabular}
\end{table}

From the posterior distribution, we find that a gap is indeed required
to fit the data, i.e. with a well defined mean radius of $73\pm3$~au
and a fractional depth that is significantly larger than zero
($0.66\pm0.15$). Although its marginalised distribution peaks at
$\sim0.7$ (see Figure \ref{fig:corner}), the gap is still consistent
with being completely devoid of dust and planetesimals, which was
ruled out for HD~107146 \citep{Marino2018hd107}. Moreover, the gap
FWHM is constrained to be $20^{+10}_{-8}$~au, which despite its large
uncertainty, is $2.4\sigma$ smaller than for HD~107146.

Because the gap's width if carved by a planet in situ is roughly equal
to the chaotic zone \citep[e.g. see Figure 5 in][]{Marino2018hd107},
which scales with the planet mass as $\Mp^{2/7}$ \citep{Wisdom1980},
then if this is the origin of the gap the planet mass must be
constrained. However, given the large uncertainty on $\wgap$, that
mass is only limited to be lower than $\sim5$~\Mjup\ ($3\sigma$). By
comparing new N-body simulations tailored to HD~92945 with the
observed visibilities (as in \cite{Marino2018hd107}), we obtain
tighter upper limits for the maximum planet mass that could carve this
gap. Depending on whether we consider particles in the planet
co-rotation zone or not, we find $3\sigma$ upper limits of
0.6~\Mjup\ and 10~\Me, respectively. Removing particles in the
co-rotation zone makes the gap deeper ($\dgap\sim1$); and therefore
the gap must be narrower and planet masses lower (see
Figure~\ref{fig:corner}). The uncertainty in the system age
(100-300~Myr) is not an important factor in this upper limit, as
planets with masses $\gtrsim 10$~\Me\ can open a gap in less than
10~Myr \citep[][]{Morrison2015, Marino2018hd107}.

Regarding the disc vertical distribution, we find that $h$ is
constrained to be $0.049\pm0.017$, although still marginally
consistent with zero. For a disc with an intermediate inclination and
a gap, the disc scale height affects how sharp the gap is along the
minor axis compared to the major axis
\citep[e.g. HL~Tau,][]{Pinte2016}. Therefore, the model prefers $h>0$
since the gap is likely seen smoother and narrower along the minor
axis, as well as the disc inner and outer edges appearing
smoother. Finally, we find a disc inclination of $65\degr\pm1\degr$
and a disc PA of $100\degr\pm1\degr$, which are consistent with
previous estimates from scattered light observations
\citep{Golimowski2011, Schneider2014}.

\begin{figure}
  \centering \includegraphics[trim=0.0cm 0.0cm 0.0cm 0.0cm, clip=true,
    width=1.0\columnwidth]{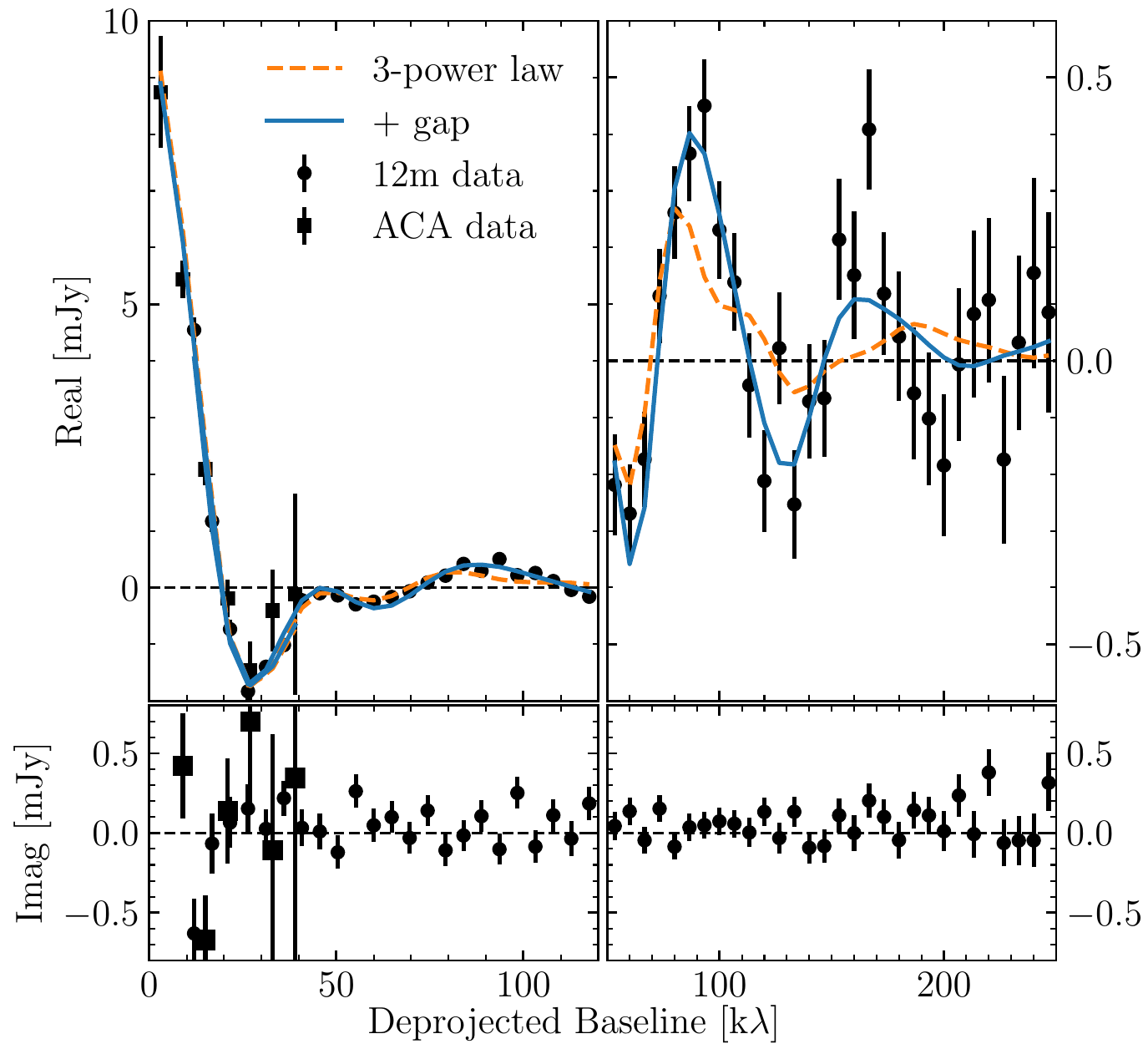}
  \caption{Deprojected and binned visibilities assuming a disc
    position angle of $100\degr$ and inclination of 65$\degr$. The
    real and imaginary components of the observed visibilities are
    presented as black error bars in the top and bottom panels,
    respectively. The errorbars represent the binned data with their
    uncertainty estimated as the standard deviation in each bin
    divided by the square root of the number of independent points. A
    continuous blue and dashed orange lines represent the triple
    power-law best-fitting model with and without a gap,
    respectively. Note that the scale in the left-hand and right-hand
    panels is different and that the x-axis is shifted. For better
    display we have not shown data points beyond 250 k$\lambda$ since
    they are all consistent with zero.}
 \label{fig:vis}
\end{figure}

At this point it is worth looking for evidence for the gap in
visibility space. In order to do this, we deproject (using the best
fit inclination and PA), radially bin and azimuthally average the
observed visibilities, and compare them with the simulated
visibilities of two models in Figure \ref{fig:vis}. These correspond
to two MCMC best fits using a 3-power law model that does not include
a gap (orange) and with the addition of a gap as discussed above
(blue). We find that between 50 and 200 k$\lambda$ there are
significant differences between the real components of the data and
the model without a gap (see top right panel in
Figure~\ref{fig:vis}). The data displays amplitudes that are larger
than the model by a factor of $\sim2$. The model with a gap, on the
other hand, is able to reproduce these amplitudes within the noise
level providing a better fit. In fact, the addition of the gap
improves the $\chi^2$ by 19, which considering the difference of three
in the total number of free parameters, is a significant
difference. Therefore, based on the gap's depth being significantly
larger than zero, evidence of the gap in the deprojected visibilities
themselves, and previous HST images which provided evidence for a
depression in the dust distribution at the same distance, we conclude
that there must be a gap or depression in the dust and planetesimal
distribution.

Despite allowing the disc to be eccentric, the marginalised
distribution of $\ed$ peaks at zero, with an upper limit of 0.1
($2\sigma$). This high upper limit is in part due to the unconstrained
stellar flux and position in these observations. We find that higher
stellar fluxes require lower eccentricities (see
Figure~\ref{fig:corner}). The disc eccentricity and disc offsets are
also correlated, such that the disc geometric center is always
centred, while the star is offset instead, providing a better fit. We
also find that $\omega$ is more likely to be near $0\degr$ than
$180\degr$ (i.e. pericentre aligned with the disc PA) when larger
eccentricities are allowed. This preference in pericentre orientation
is possibly due to the west side of the disc appearing more extended
(see Figures \ref{fig:alma} and \ref{fig:radial_profile}).


\begin{figure*}
  \centering
 \includegraphics[trim=0.4cm 0.4cm 0.4cm 0.3cm, clip=true, width=1.0\textwidth]{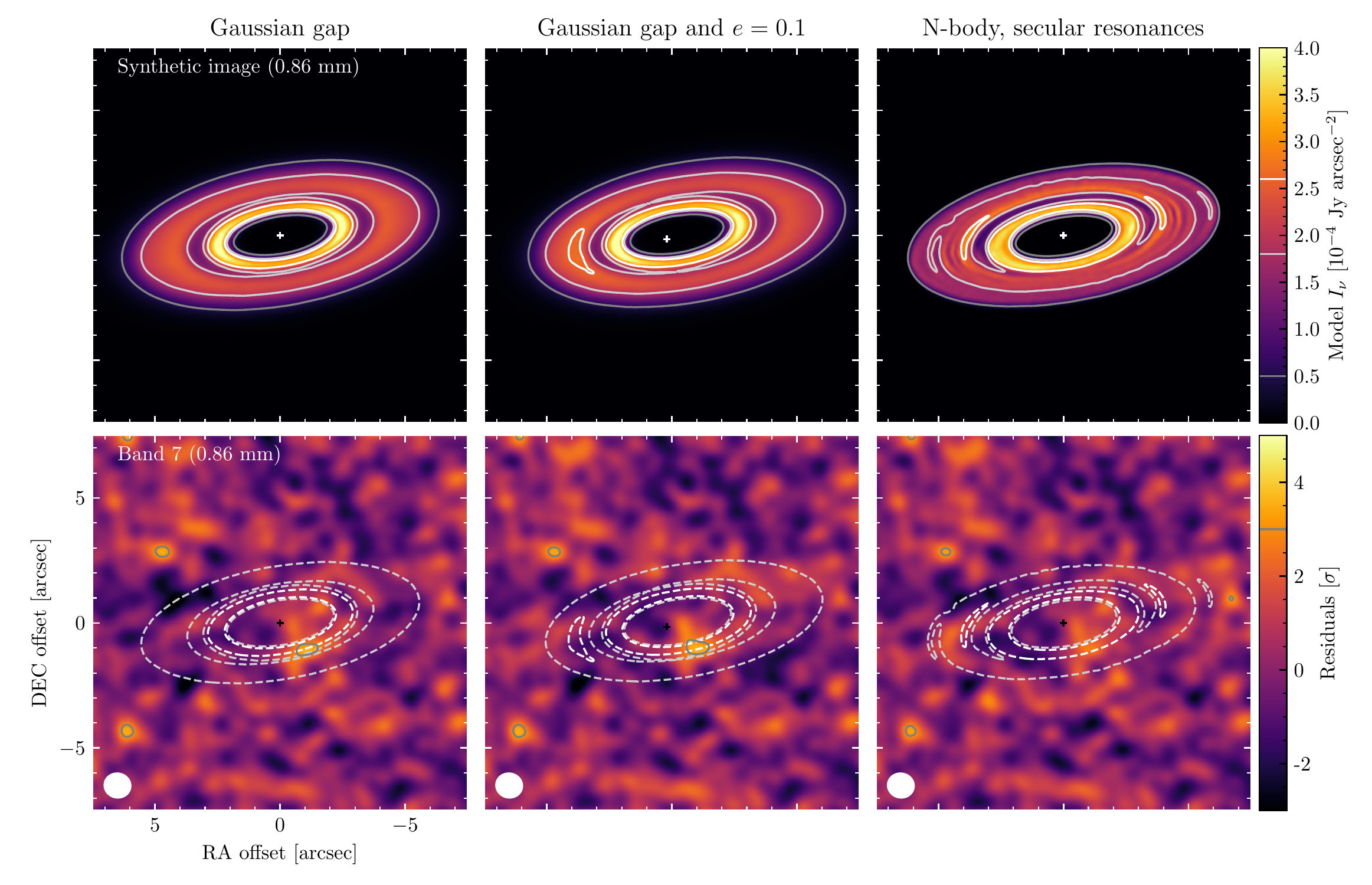}
 \caption{Best fit model images (top) and dirty map of residuals
   (bottom) at 0.86~mm. \emph{\textbf{Left column:}} 3-power law model
   with a Gaussian gap and zero eccentricity. \emph{\textbf{Middle
       column:}} 3-power law model with a Gaussian gap and a disc
   global eccentricity of 0.1 with a pericentre at the disc
   PA. \emph{\textbf{Right column:}} Secular resonance model based on
   N-body simulations. The residuals are computed using natural
   weights and with an outer taper of $0\farcs7$, leading to a
   synthesised beam of $0\farcs88\times0\farcs80$ and an rms of
   20~$\mu$Jy~beam$^{-1}$, respectively. The solid contours in the
   lower panels represent 3$\sigma$ levels, while the dashed contours
   correspond to the model images in the top panels.}
 \label{fig:res}
\end{figure*}

In Figure \ref{fig:res} we present best fit model images (top) and the
corresponding residuals (bottom) for a disc with zero eccentricity and
one with 0.1 eccentricity with pericentre at the South East ansa. In
the residual images for both models there are a few 70~\uJybeam
($\sim3\sigma$) peaks, one of them located towards the South West
($3.5\sigma$ in the circular disc model) where we found that the disc
appears brighter and the gap seems absent. Whether this observed
residual is due to a true disc asymmetry or background emission is
uncertain. Sub-millimetre galaxies are ubiquitous at these frequencies
in deep observations. The expected number of sub-millimetre galaxies
brighter than 70$\mu$Jy at 0.88~mm within a square region of
$15\arcsec\times15\arcsec$ is 8 \citep{Simpson2015}, therefore it is
not surprising that we find between four and five 70~\uJybeam peaks in
the residual images. Moreover, we expect to find about 2
sub-millimetre galaxies brighter than 70~$\mu$Jy in the area covered
by the disc in the sky, therefore it is possible that the observed
\textit{clump} is a background object. To rule out this clump being a
background object new observations in 2020 or later with a similar
sensitivity and resolution could confirm if it is co-moving or not
with HD~92945 given its stellar proper motion
\citep[$|\mu|=$220~mas~yr$^{-1}$,][]{Gaiadr2}. Alternatively,
observations at a different wavelength could be used to measure the
disc and clump spectral index and see if they differ, which could
suggest that the observed residual is background emission. The
difference in spectral index was used as an evidence for the
extra-galactic origin of a clump in HD~95086 ALMA observations
\citep{Su2017, Booth2018}.



\section{Gap opening by secular resonances}
\label{sec:SR}

Given the putative asymmetry found in the ALMA data, we begin this
section by describing a dynamical model which could explain both the
gap and asymmetry observed in the HD 92945 disc via interactions with
a two-planet system, with both planets interior to the disc. Such a
system of planets will excite the eccentricities of planetesimals in
the disc over certain ranges of semi-major axis where the
planetesimals' pericentres are precessing at close to one of the
eigenfrequencies of the system, which are the frequencies that set the
precession rates of the planets themselves. The planetesimals
satisfying this condition are said to be in secular resonance with the
planets \citep[e.g.][]{MurrayDermott1999}. \cite{Yelverton2018}
demonstrated that a gap can form in the disc at the secular resonance
locations, as the eccentric planetesimals there spend more time per
orbit close to apocentre than at a radius equal to their semi-major
axis. The resulting gap is asymmetric, with a width and depth that
varies with azimuthal angle, as the inner ring of the disc is offset
relative to the outer.

In their paper, they applied this model to HD~107146, finding that it
is difficult to produce a gap that is both as wide and as axisymmetric
as in the observations of \cite{Marino2018hd107}. The gap in HD~92945
is located at around the same radial distance as that in HD~107146,
but is narrower and does not appear to be axisymmetric; the stars also
have similar estimated ages. Thus, here we compare the results of
their simulation A (with planetary semi-major axes of 3 and 26~au,
masses of 1.5 and 0.6~\Mjup\ respectively, and initial eccentricities
both set to 0.05) with our ALMA observations.


Note that this particular choice of planets is not unique, and certain
other configurations of planets are expected to work equally well
\citep[see for example Fig. 7 in][and Figure \ref{fig:planetsr}
  here]{Yelverton2018}. It may be possible with some fine-tuning to
find a combination of planets which better reproduces the
observations, but our aim here is simply to demonstrate that the
secular resonance model (SR model hereafter) can explain the disc's
structure at least as well as simple parametric models, rather than to
identify a ``best fit'' system of planets.

Based on the output of simulation A, we simulate observations that we
compare directly to the data. To do this, we take the orbital elements
of each simulated particle ($\sim3500$ particles between 45 and
130~au) at the end of the simulation (100~Myr) and populate their
orbits with 300 points with randomly distributed mean anomalies. We
impose an initial power law surface density proportional to
$r^{\gamma}$ by weighting each particle according to their initial
semi-major axis. This weighting scheme allows us to leave the surface
density slope as a free parameter, similar to the parametric model in
\S\ref{sec:parmodel}. We translate the resulting density distribution
to a dust density field. Finally, we produce synthetic images using
RADMC-3D, which we use to compare with observations in visibility
space. In order to obtain the best match that we can, we vary some
disc parameters such as the total dust mass, $\gamma$, the minimum and
maximum initial semi-major axes since the HD~92945 disc is somewhat
narrower than that of HD~107146, and the disc orientation on the
sky. In addition, we linearly scale the semi-major axis of all
particles such that the gap radial location best matches the
observations, although we found a best match by only scaling the
semi-major axes by 0.99. Therefore, in total we vary 11 parameters:
$\Md$, $\rmin$, $\rmax$, $\gamma$, $i$, PA, four parameters to fit any
offset in the 12m and ACA data, and one to scale the simulation.

In the right panels of Figure \ref{fig:res} we present a model image
(top) and residuals after subtracting the model visibilities from the
observations (bottom). The model image shows a variety of features
such as bright narrow rings, spiral features and a gap that is
asymmetric as described above. We oriented the disc such that the gap
is narrowest towards $\omega\sim100\degr$ (PA$\sim200$). We find that
residuals of the SR model are similar compared to the previous
parametric models, although no emission above $3\sigma$ is present in
these residuals near the disc. Moreover, we find a $\chi^2$ that is
larger only by 11 compared to the circular disc (with
$2.53\times10^{6}$ data points and a total of 11 free parameters vs 19
for the eccentric disc model presented in \S\ref{sec:parmodel}). This
small difference in $\chi^2$ therefore favours the SR model when
compared against the full eccentric disc model presented in
\S\ref{sec:parmodel} because it has fewer parameters (based on a
$\chi^2$ difference test and the Bayesian Information Criterion).

\begin{figure}
  \centering \includegraphics[trim=0.0cm 0.0cm 0.0cm 0.0cm, clip=true,
    width=1.0\columnwidth]{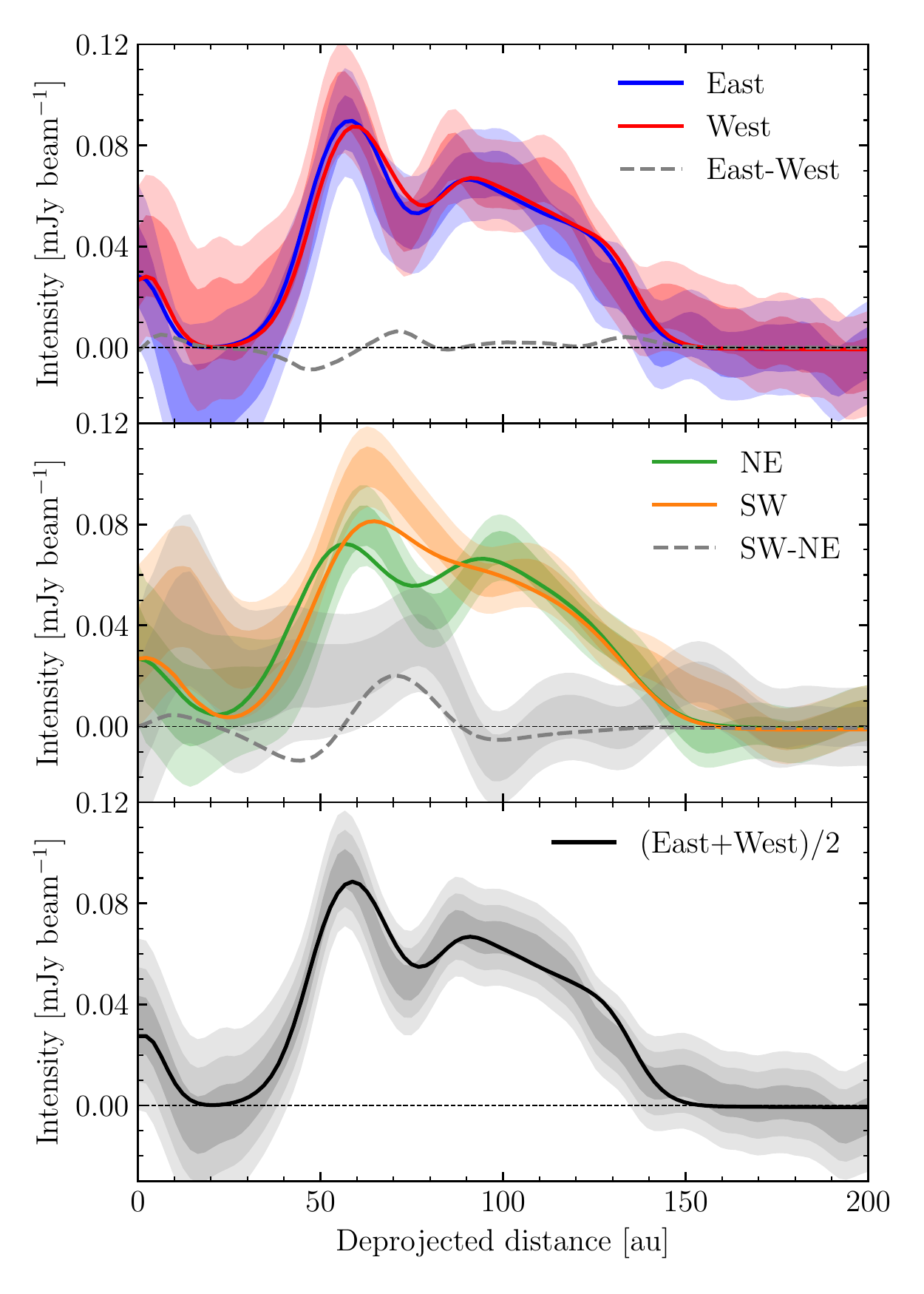}
 \caption{Deprojected surface brightness profile of the SR model,
   computed by azimuthally averaging the emission from simulated
   observations (Clean image with natural weights) within $60\degr$
   wide wedges. \textbf{\textit{Top:}} Comparison between the East
   (blue) and West (red) side of the disc along the disc major axis,
   i.e. at a mean PA of $100\degr$ and $280\degr$,
   respectively. \textbf{\textit{Middle:}} Comparison between opposite
   wedges with a mean PA of $45\degr$ (NE, green) and $225\degr$ (SW,
   orange). \textbf{\textit{Bottom:}} Average between the East and
   West sides. The grey dashed lines represent the difference between
   the East and West wedges (top) and SW and NE wedges (middle). The
   shaded regions in the top and middle panels correspond to the 1 and
   $2\sigma$ confidence regions of the observations, while in the
   bottom, panel they represent 1, 2 and $3\sigma$ confidence
   regions.}
 \label{fig:mrp}
\end{figure}

We also compare the SR model with the observations by computing radial
profiles as in \S\ref{sec:continuum}. Figure \ref{fig:mrp} compares
the intensity radial profiles obtained from simulated observations
using the same uv sampling, but without any noise (continuous and
dashed lines) with the radial profiles observed from the observations
(shaded regions). Similar to HD~92945, the East and West sides of the
disc are symmetric, or at least consistent within the noise in the
original data, which is also reproduced with the parametric model with
zero eccentricity. The observed gap width and depth are also
reproduced by the SR model. Moreover, the SR model has an asymmetry
between the NE and SW sides of the disc, which appears similar to the
observations. In particular, the SW side is brighter than its opposite
at a deprojected distance of 70-80~au (i.e. within the gap), as we
found in our observations. Whether this is a true asymmetry or due to
a background object is uncertain.

In conclusion, the SR model is able to reproduce the observations
suggesting that a planet on a wide orbit at $\sim73$~au is not the
only plausible scenario to explain the observed gap. New observations
could confirm the disc asymmetry, which with a higher S/N would make
it possible to better quantify the width a depth of the gap.

\section{Discussion}
\label{sec:dis}

In this section we discuss the potential origins of the gap in the
planetesimal disc around HD~92945, and how the new ALMA observations
provide constraints to these models or rule them out.

\subsection{The gap origin}

\subsubsection{Planet in situ}
\label{sec:pinsitu}
Thanks to instruments such as ALMA, SPHERE and GPI, in the last few
years multiple gaps have been discovered in nearby circumstellar
discs. Most of them in protoplanetary discs either in dust continuum
or gas emission \citep{HLtau, Andrews2016, Tsukagoshi2016, Isella2016,
  Thalman2016, Cieza2017, vanderPlas2017, Sheehan2018, Teague2018,
  Avenhaus2018, Fedele2018, Clarke2018}, suggesting the presence of
planets at tens of au. As shown here and in \cite{Marino2018hd107},
gaps are also observed in some wide debris discs in the distribution
of planetesimals traced by ALMA observations. A natural question then
arises which is \textit{are these gaps in planetesimal discs
  primordial or secondary?} i.e. were these gaps present in the solid
distribution that formed these planetesimal discs or were they carved
after planetesimal formation?

Let us assume that the gap was created by an ice giant planet, formed
in situ early during the protoplanetary disc lifetime (e.g. through
pebble or planetesimal accretion), long before the observed
planetesimal disc was formed. Such a planet would have created a gap
or at least a depression in the dust distribution due to the perturbed
gas pressure profile \citep[e.g.][]{Rice2006, Zhu2012, Owen2014,
  Rosotti2016}. Therefore, planetesimal formation could have been
hindered near the planet orbit after it formed, creating a
\textit{primordial} gap. On the other hand, if the planet attained its
final mass after disc dispersal (e.g. through planetesimal accretion),
or if the planet formed from the same distribution of planetesimals
that formed the debris disc, then the gap around its orbit should be
set by the size of its chaotic zone, where the orbits of minor bodies
become unstable and are scattered away on short timescales. In this
latter case, the gap would be of \textit{secondary} origin.

\begin{figure}
  \centering \includegraphics[trim=0.0cm 0.0cm 0.0cm 0.0cm, clip=true,
    width=1.0\columnwidth]{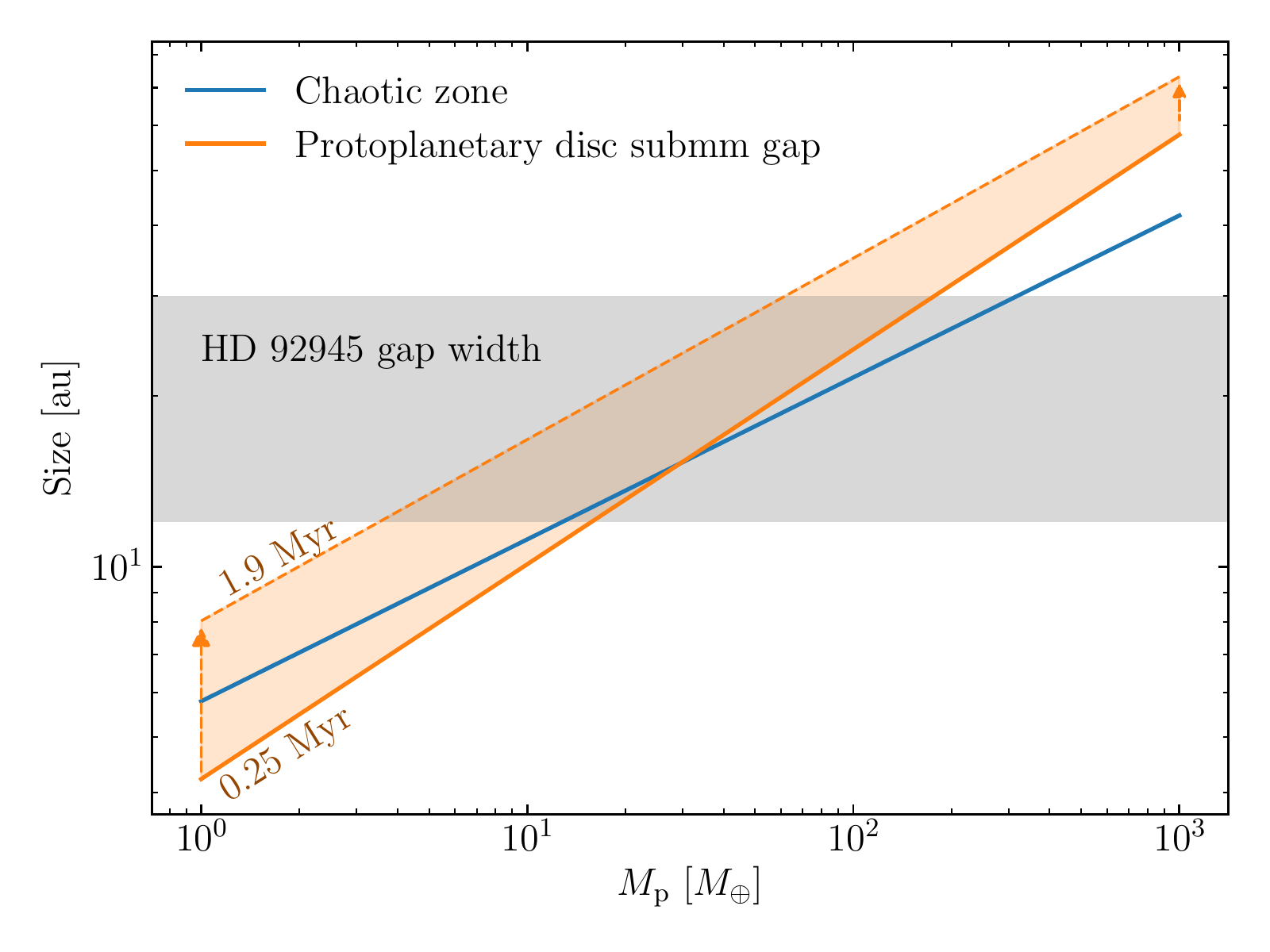}
 \caption{Comparison between the size of the chaotic zone
   ($\ap\pm1.5\ap(\Mp/\Mstar)^{2/7}$, blue) and a gap in the dust continuum
   emission \citep{Rosotti2016} after 0.25 Myr (continuous orange
   line) and 1.9 Myr (dashed orange line) as a function of planet mass
   for a planet at 73~au. The horizontal dashed region represents the
   constraints on the gap FWHM derived in \ref{sec:parmodel}.}
 \label{fig:gapsize}
\end{figure}

In Figure \ref{fig:gapsize} we compare the size of the chaotic zone
(blue) and the gap produced in the distribution of mm-sized dust in a
protoplanetary disc as a function of planet mass (orange lines), based
on numerical simulations by \cite{Rosotti2016}. In
\cite{Rosotti2016}'s simulations, a planet creates a gap in the gas
distribution or at least a pressure maximum outside its orbit in the
protoplanetary disc, affecting the dust grains and producing a
depression or gap in the dust distribution around its orbit. The
protoplanetary disc gap after 400 orbits (0.25 Myr of evolution,
orange continuous line) has a very similar size to the chaotic zone
(they differ by $\lesssim20\%$ for planet masses between 1~\Me\ to
1~\Mjup), but after 3000 orbits (1.9 Myr) the width of the
protoplanetary disc gap has converged and is 50-80\% larger than the
chaotic zone. Therefore, we expect primordial gaps in debris discs to
be similar or even larger compared to secondary gaps for the same
planet mass.


Understanding whether a gap is primordial or secondary is important if
we want to infer the mass of a gap clearing planet from debris disc
observations. For example, assuming the gap is primordial, the planet
mass is only constrained between 3 and 100~\Me\ ($1\sigma$ see Figure
\ref{fig:gapsize}), while if secondary, between 10 to 200 \Me. A
potential way to discern between these two origins is to look for how
much material is left within the gap. If the planet is massive enough,
in the primordial scenario we expect no dust and planetesimals within
the gap, since no dust was able to filter through the gas pressure
maxima created outside the planet orbit in the protoplanetary
disc. Moreover, planetesimals would potentially only form beyond the
chaotic zone according to Figure \ref{fig:gapsize}, and would not
undergo scattering with the planet as in the secondary scenario. On
the other hand, if the gap is of secondary origin and cleared through
scattering encounters, some material might still be present inside the
gap from solids on highly eccentric orbits that have not been ejected
yet depending on the scattering timescales
\citep[e.g.][]{Tremaine1993}.

A caveat to the comparison above is that we are assuming planets have
no significant migration. Planet migration through a gaseous disc
could significantly change how the dust is distributed, possibly
opening an even wider primordial gap if the planet migrates
outwards. On the other hand, if the gap is carved through scattering
of planetesimals, we expect that those interactions will lead to a
change in the planet's angular momentum and thus semi-major axis
\citep{Fernandez1984, Ida2000, Gomes2004, Kirsh2009,
  Ormel2012}. Planet migration could open a gap significantly wider
than the planet's chaotic zone \citep[e.g.]{Morrison2018}, and
moreover, it could lead to an asymmetry between the inner and outer
edge of the gap \citep[e.g.][]{Marino2018hd107}.

Planetesimal driven migration of a single planet tends to be inwards
\citep{Kirsh2009} and only occurs at a significant level when the disc
mass within a few Hill radii from the planet's orbit is similar or
larger than the planet mass. In order to compute the primordial mass
surface density at around 73~au, we use the estimated surface density
of millimetre dust for HD~92945 and Equation 7 in \cite[which had a
  typo where the factor preceding the right hand side should be
  $2\times10^{-7}$, and here we correct][]{Marino201761vir}
\begin{equation}
  \begin{split}
    \Sigma_{\rm mm}(r>r_c) \approx 2\times10^{-7} \left(\frac{r}{1\ {\rm AU}}\right)^{0.6\alpha+0.9}
    \left(\frac{t_{\rm age}}{1\ \mathrm{Gyr}}\right)^{-0.4}\\
    \left(\frac{D_{\max}}{100\ \mathrm{km}}\right)^{-0.1}
    \left(\frac{\Sigma_0}{1\ \mathrm{MMSN}}\right)^{0.6} 
    {\rm M}_{\oplus}\ {\rm AU}^{-2}, \label{eq:Sigmamm}
  \end{split}
\end{equation}
where $\alpha$ is the mass surface density profile (including
planetesimals), $t_{\rm age}$ is the age of the system, $D_{\max}$ is
the maximum planetesimal size and $\Sigma_0$ is the primordial surface
density at 1~au in units of the Minimum Mass Solar Nebula
\citep[MMSN,][]{Weidenschilling1977mmsn,Hayashi1981}. We estimate the
initial surface density of solids at 73~au to be
0.005-0.01~\Me~au$^{-2}$ (depending on the age), which integrated
between 50 to 130 au results in a total disc mass of 100-200~\Me\ (for
an age of 100-300~Myr, respectively)\footnote{Assuming a radial
  mass distribution proportional to $r^{-1.5}$, consistent with the
  derived slope of millimetre grains and Equation 6 in
  \cite{Marino201761vir}.}. This means that for an ice giant planet of
$\lesssim20$~\Me\ there would have been enough solid mass around its
orbit to drive migration.

Current constraints in the presence of planets at $\sim70$~au by the
Gemini/NICI planet-finding campaign are not very constraining, only
discarding planets more massive than
$\sim9$~\Mjup\ \citep{Biller2013}. Future direct imaging observations
with new generation instruments such as JWST/NIRCam or MIRI
\citep[e.g.][]{Beichman2010}, or with ground base 30m class telescopes
such as E-ELT/METIS \citep{Quanz2015} or TMT/MICHI \citep{Packham2018}
could detect planets with masses much lower than current upper limits
and that are consistent with the measured gap width and depth,
i.e. below 10~\Mjup.

A potential problem in the planet in situ scenario is that these
observations also showed an asymmetry in the disc when comparing the
North East and South West sides. Strong asymmetries are not expected
if the planet is on a circular orbit. Hence, if the asymmetry is real
and not due to a background object (e.g. sub-millimetre galaxy), then
this would indicate that the gap has a different origin. Moreover, ice
giant planet formation via core accretion at these distances is
challenging according to various planet formation models
\citep[e.g.][]{Safronov1969a, Pollack1996, Kenyon2008, Bitsch2015},
although these remain to be tested at these distances.

\subsubsection{Inner planets}

Because planets might be more likely to be found at smaller radii, it
is worth considering scenarios in which an inner planet with a
semi-major axis ($\ap$) within 50~au could produce a gap at around
73~au. In protoplanetary discs, multiple gaps in the gas and dust
distribution could be produced by multiple spiral arms excited by a
single planet \citep{Dong2017multiplegaps, Bae2017}. However, in those
models the most external of these gaps is still too close to the
planet orbit to explain our observations that suggest
$\rgap/\ap\gtrsim1.5$ (assuming it is caused by an inner planet
interior to $\rmin$).

A secondary origin gap could also be produced by inner planets. Four
scenarios have been investigated so far: mean motion resonances
\citep[MMRs,][]{Tabeshian2016, Regaly2018}, resonant trapping by a
migrating planet \citep{Wyatt2003, Wyatt2006}, secular interactions
\citep{Pearce2015doublering} and secular resonances
\citep{Yelverton2018}. In the first case, a massive planet can create
both interior and exterior gaps, especially at the 3:2 MMR. An
interior planet at $\sim45$~au on a low eccentricity orbit
($\ep\lesssim0.05$) would open a gap at the right location
($\sim73$~au). Such a gap would be asymmetric, with a width that
varies as a function of azimuth similar to our observations.  In order
to explain the observed width ($20^{+10}_{-8}$~au), a 6-19~\Mjup\ mass
planet would be required \citep{Tabeshian2016}, which is just below
the most conservative direct imaging upper limit of $\sim9$~\Mjup\
\citep[assuming an age of 100~Myr][]{Biller2013}. However, such a high
planet mass means that the planet would push the inner edge of the
disc out to a radius $\gtrsim60$~au (outer edge of its chaotic zone),
which is inconsistent with our observations. Therefore, we rule out
this scenario.

In the second scenario, a planet migrated outwards through a disc
trapping planetesimals in mean motion resonances and eccentric
orbits. As shown by \cite{Wyatt2006}, small dust generated from
collisions of planetesimals trapped in the 2:1 MMR exit the resonance
due to radiation pressure, creating an axisymmetric wide disc with a
semi-filled gap, resembling the double peaked profile seen in
HD~92945's HST observations \citep{Golimowski2011}. Such a structure
would be absent at longer wavelengths that trace larger grains;
however, our ALMA observations show that the gap is also present in
the distribution of millimetre-sized grains. This means that the
planet that migrated must be less massive than
$\sim10^{-2}$~\Me\ \citep[Equation 14 in][]{Wyatt2006}, which would
make the resonant population unlikely\footnote{The planet would need
  to migrate at $10^{-4}$~au~Myr$^{-1}$, too slow to sweep and capture
  a significant number of planetesimals}. Alternatively, if the
migration is stochastic, planetesimals might be only weakly bound to
the resonance, making it easier for larger grains to exit and form an
axisymmetric structure.

In the third scenario, a planet on an initially eccentric orbit
(e.g. after a scattering event with an inner planet) interacts with a
planetesimal disc of similar mass, opening a gap through secular
interactions that is asymmetric. During the evolution multiple
scattering events damp the planet's eccentricity and its orbit is
circularised, ending up near the inner edge of the disc on an orbit
close to where the surface density peaks interior to the gap. This
scenario could produce a gap similar to the one observed as well as
the possible asymmetry since the gap width varies with azimuth in that
model. The inner planet must be of a similar mass compared to the
disc, i.e. $\sim100$~\Me\ based on disc mass estimates discussed in
\S\ref{sec:pinsitu}. However, \cite{Pearce2015doublering} showed that
the resulting inner edge of the disc is not steep, but rather smooth
as a result of planet disc interactions. This is in contradiction with
our observations that suggest a very steep inner edge, although it is
not clear if the addition of an inner planet closer in (that scattered
the outer planet) in \cite{Pearce2015doublering} simulations could
lead to a steeper inner edge.

\begin{figure}
  \centering \includegraphics[trim=1.0cm 0.5cm 1.0cm 0.5cm, clip=true,
    width=1.0\columnwidth]{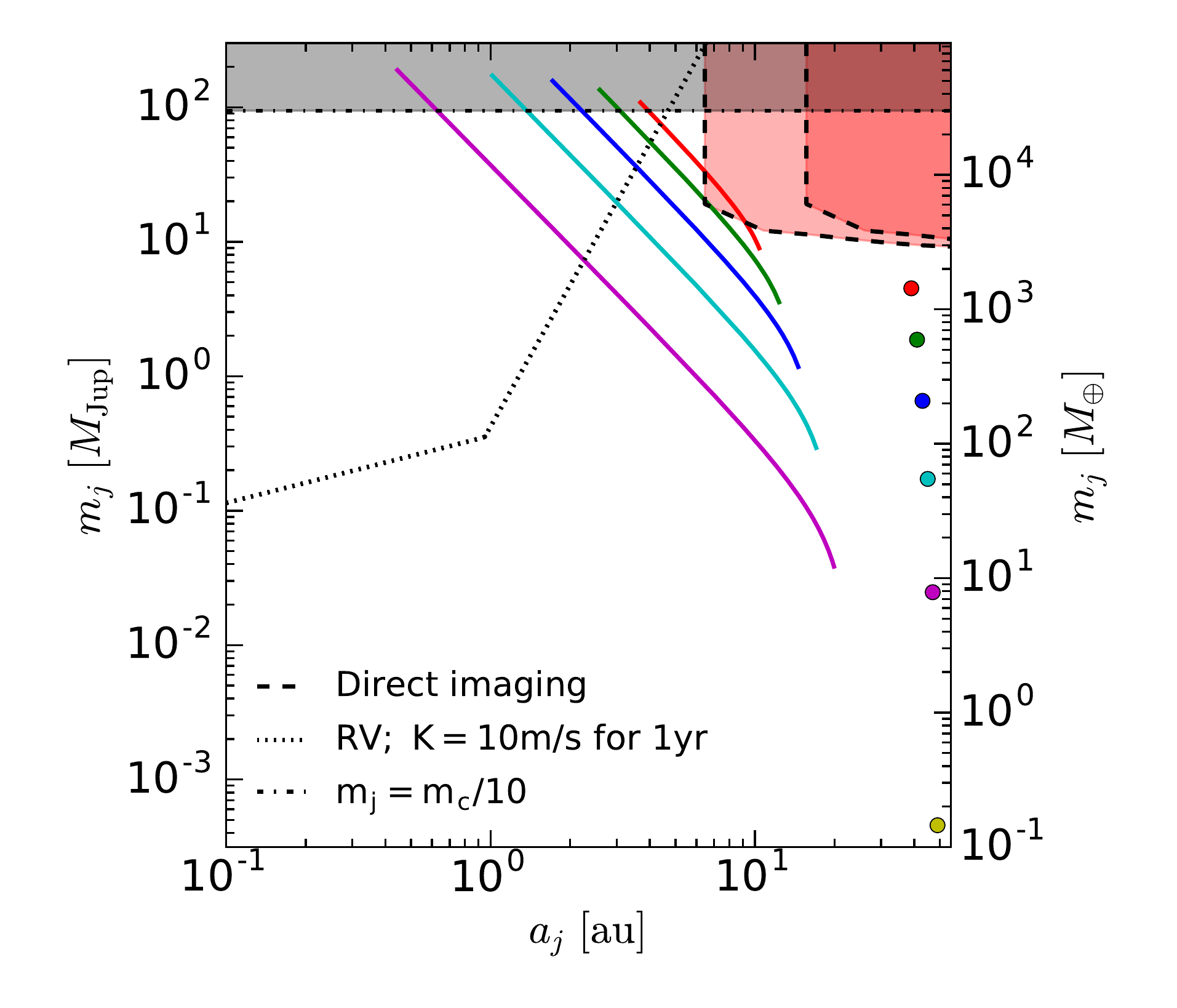}
 \caption{Mass and semi-major axes of two putative planets interior to
   HD~92945 disc and that produced the observed gap. The outer planet
   location is represented by coloured circles chosen such that it
   truncates the disc at 50~au (observed inner edge). The mass and
   semi-major axis of the corresponding innermost planet are
   represented by continuous lines, that ensure that the gap is at
   73~au, it is set in less than 300~Myr and the two planets are on
   stable orbits. With the outer planet at 49~au, the secular
   timescale at 73~au exceeds 300~Myr for any choice of inner planet,
   so there is no corresponding yellow line shown. This figure is
   analogous to Figure 7 in \protect\cite{Yelverton2018}. The red
   shaded regions represent planet masses excluded by direct imaging
   \citep[][assuming that the planet is near the minor or major axis
     of the disc]{Biller2013}, while the grey region in the top
   excludes companion masses that exceed one-tenth of the central
   star. The region above the dotted line is where radial velocity
   measurements could easily detect a planet with a precision of
   10~m~s$^{-1}$ and spanning one year, although there are no
   published RV results.}
 \label{fig:planetsr}
\end{figure}

Finally, in \S\ref{sec:SR} we showed how two inner planets could open
a gap within the planetesimal disc through secular resonances. Figure
\ref{fig:planetsr} summarises the constraints on these planets
assuming that the outermost planet truncated the disc at 50~au, the
gap is located at 73~au, and the gap is set in less than 300~Myr
(upper limit for the age of the system). The outermost planet must
have a semi-major axis of about around 40-45~au depending on its mass,
which would be constrained to be larger than 10~\Me\ \citep[otherwise
  the timescale to create the gap is longer than
  300~Myr,][]{Yelverton2018}. Its mass, on the other hand, must be
lower than direct imaging upper limits of
$\sim9$~\Mjup\ \citep[derived by][and represented with a red shaded
  region]{Biller2013}. These upper limits only apply to the innermost
planet if it has a semi-major axis larger than $\sim7$~au. Its
semi-major axis also cannot be larger than 20~au to ensure dynamical
(considering that the planets cannot be closer than 5 mutual Hill
radii). There is no current lower limit for its semi-major
axis. However, if it was smaller than 2~au, it would be readily
detectable using radial velocities (dotted line), although we are not
aware of any published radial velocity detection or upper limit even
though it has been monitored, therefore planets could still lie in
that region.  Note that the location of the gap and timescales depend
on the planet mass and semi-major axis ratios, therefore the exact
planet masses and semi-major axes required to explain the observations
are degenerate. As commented before in \S\ref{sec:pinsitu} new
generation instruments in JWST or 30m class telescopes could detect
planets beyond 10 au with masses significantly lower than current
direct imaging limits, potentially confirming or ruling out this
scenario.

\subsubsection{No planet scenarios}

Multiple alternative scenarios have been proposed to explain gaps in
protoplanetary discs that do not require the presence of planets. Some
invoke changes in the dust opacities near condensation fronts (snow
lines) due to changes in grain growth efficiency \citep{Zhang2015,
  Okuzumi2016}, or gas dust interactions \citep{Takahashi2014,
  Ruge2016, Flock2017, Dullemond2018}. We do not expect that local
decrements in the dust opacities due to dust growth in protoplanetary
discs to be imprinted in the radial distribution of dust produced by
planetesimal grinding, unless planetesimal properties are different
within these gaps (e.g. strength, size distribution, formation
efficiency). On the other hand, gas dust interactions can affect the
solid mass distribution and thus the planetesimal distribution
inherits this gap. Therefore, the latter is a potential explanation
for the observed gaps in debris discs too.

\subsection{Volatile delivery}

If the gap was of secondary origin and cleared by a planet on an orbit
at 73~au, some of the planetesimals would get scattered inwards where
they could deliver volatiles such as water to inner planets that
potentially formed dry \citep[e.g. like
  Earth,][]{Morbidelli2000}. Although a single planet on a circular
orbit cannot scatter material into the temperate regions where
habitable planet may lie in this system \citep{Bonsor2012analytic,
  Bonsor2012nbody}, additional inner planets can facilitate this
process as long as the orbits of the two outermost planets are close
enough. This is because the pericentre of particles interacting with a
single planet is constrained to be larger than $q_{\min}$, which
depends on the planet mass and initial Tisserand parameter \citep[or
  orbital eccentricity and inclination,][]{Tisserand1896}. If the
second outermost planet has a semi-major axis smaller than $q_{\min}$,
particles will never get scattered by the second outermost planet and
inward scattering will be hindered.

Although the planet in situ scenario does not require the presence of
additional inner planets, these cannot be ruled out either. In order
to assess whether inward scattering is possible, here we consider the
best case scenario: the second planet is as far out as possible
without disrupting the disc, i.e. on an orbit at
$\sim50$~au. Therefore one of the most basic conditions for inward
scattering and volatile delivery from 73~au in HD~92945 is that the
$q_{\min}<50$~au, such that planetesimals can get scattered by the
second planet. Using Equation 4 in \cite{Marino2018scat} and assuming
that planetesimals have zero eccentricities and inclinations, we
estimate that only if the planet at 73~au is more massive than 7~\Me\,
could volatile rich planetesimals reach 50~au and be scattered further
in. For higher initial eccentricities and inclinations the minimum
planet mass is lower, e.g. with initial eccentricities of 0.05 and
inclinations of $\sim1\degr$ the minimum planet mass is 2~\Me. For
HD~107146 on the other hand, the same analysis leads to a minimum
planet mass of $\sim30$~\Me\ assuming zero initial eccentricities and
inclinations since the inner radius is at 40 rather than 50~au. Larger
initial eccentricities of 0.05 and inclinations of $\sim1\degr$
translate into a lower minimum mass of 20~\Me. These planet masses for
HD~107146 are at the limit since planet masses much higher than
$\sim30$~\Me\ were found to be inconsistent with the width and depth
of the gap \citep{Marino2018hd107}. We conclude therefore that the
material cleared from the gaps in HD~107146 and HD~92945 could have
been delivered to inner planets if: the gap is of secondary origin and
cleared by a planet in situ; a second outer planet lies near the inner
radius of the disc; and the planet within the gap is more massive than
7~\Me\ in HD~92945 and 30~\Me\ in HD~107146 (assuming very low initial
eccentricities and inclinations).

\subsection{Gas mass upper limits}
\label{sec:gasmass}


Here we derive gas mass upper limits based on the flux limits derived
in \S\ref{sec:gas} for CO (3-2) and HCN (4-3) line emission. Given the
low gas densities expected in the disc, non-LTE effects can be
significant and conversion from flux to gas mass is highly dependent
on the density of collisional partners and gas kinetic temperature,
which remain largely unknown \citep{Matra2015}. We use the code
developed by \cite{Matra2018} to compute CO gas masses given the flux
upper limit, a wide density range of collisional partners (to cover
from radiation dominated to LTE regimes), a range of kinetic
temperatures, and taking into account the cosmic microwave background,
stellar and dust thermal radiation. We find a CO gas mass upper limit
of $3\times10^{-5}$~\Me\ in the radiation dominated regime, which is
orders of magnitude greater than $3\times10^{-7}$~\Me\ if CO were in
LTE. For HCN there is no equivalent non-LTE tool to calculate line
populations, hence we simply assume LTE. Assuming a kinetic
temperature of 40~K (worst case scenario), we find a HCN gas upper
limit of $10^{-9}$~\Me.

Assuming the most conservative upper limit of $3\times10^{-5}$~\Me\ of
CO and that primordial gas could still be present (with a CO to H$_2$
ratio of 10$^{-4}$), we find an upper limit on the CO column density
of $\sim10^{17}$~cm$^{-2}$ and $\sim10^{21}$~cm$^{-2}$. This is only
enough to shield CO molecules by a factor of 660 \citep{Visser2009},
resulting in a photodissociation timescale of 80~kyr, i.e. too short
for CO to survive for the age of the system (100-300~Myr). Note that
such a high H$_2$ column density is inconsistent with CO being in
non-LTE \citep[see Figure 4 in][]{Matra2015}, hence the
photodissociation timescale of CO must be even shorter.

Therefore, since either CO or HCN gas present in the system would need
to be of secondary origin (i.e. released from solids in the
collisional cascade), we can link the upper limits above with the
volatile composition of planetesimals \citep[e.g.][]{Zuckerman2012,
  Marino2016, Matra2017betapic, Matra2017fomalhaut}. We find that the
CO non-detection is not surprising as even with an extreme 99\% CO
mass fraction in planetesimals, the mass loss rate in the HD~92945
debris disc is not high enough to maintain detectable levels of CO
\citep[using equation 2 in][]{Matra2017fomalhaut}. For HCN, we find a
mass fraction upper limit of 3\% in planetesimals, which is an order
of magnitude higher than the inferred levels in comets
\citep{Mumma2011}, therefore not constraining when assuming LTE.

\section{Conclusions}
\label{sec:conclusions}

In this paper we have presented the first ALMA observations of the
debris disc around the nearby K0 star HD~92945. After HD~107146, this
is the second system that we know of with an outer belt of
planetesimals divided by a gap. These observations reveal the emission
of mm-sized grains for which radiation forces are negligible;
therefore tracing the distribution of the parent planetesimals. We
find that the planetesimal belt emission is detected from 50 to
140~au, similar to HST images tracing the $\mu$m-sized grains pushed
out by radiation pressure. Within this broad debris disc we found that
the disc has a gap at about 73~au, which was tentatively detected with
HST. Interestingly, we found that the disc looks asymmetric, with the
South West side being brighter and with a no clear depletion or gap at
75 au. Whether this is a true disc asymmetry or due to background
emission is not clear and new observations are necessary to assess the
nature of the observed asymmetry.

In order to constrain the disc morphology we fit the data in
visibility space using a parametric disc model. We found that the disc
has steep inner and outer edges roughly at 52 and 121 au, with a close
to flat surface density of dust in between (as expected from
collisional evolution models). More importantly, we found that the gap
has a mean radius of $73\pm3$~au, a width of $20^{+10}_{-8}$~au and a
fractional depth of $0.66\pm0.15$. By comparing N-body simulations
with the data assuming that the planetesimal gap was cleared by a
planet at 73~au, we found that the planet must have a mass lower than
0.6~\Mjup\ and even lower than 10~\Me\ if no material is present in
the planet co-rotation region. We also allowed for a disc global
eccentricity, taking into account how the surface density should vary
as a function of azimuth. We found that the disc eccentricity is
consistent with zero, with a 2$\sigma$ upper limit of 0.1

Motivated by the putative disc asymmetry we explored a dynamical model
where two inner planets can open a gap through secular resonances in a
broad exterior debris disc. We showed that this scenario can explain
both qualitatively and quantitatively the observations, i.e. the gap
location, width and potential asymmetry. Using this model we obtain a
fit to the observations that is as good as with a full parametric
model which has more free parameters. The putative inner planets are
consistent with direct imaging upper limits, and represent an
alternative scenario to explain the gap through planet disc
interactions. If the disc is truly asymmetric, this would strongly
favour this scenario.

Given the ubiquity of gaps found in protoplanetary discs in recent
years, we also discussed the possibility of the gap being primordial
rather than secondary, i.e. originated from a gap in the solid/dust
distribution in a protoplanetary disc before the planetesimal disc was
formed. A primordial gap could be caused by the presence of a planet
in situ or due to gas dust interactions, shaping the distribution of
dust grains that will grow to form a planetesimal disc. We ruled out
other primordial scenarios that can open gaps at multiple radii via
spiral wakes in protoplanetary discs and snow lines. If the gap is of
secondary origin instead (i.e. the gap was carved after planetesimal
formation), scenarios invoking mean motion resonances, resonant
trapping after gas dispersal, and secular interactions between a
planetesimal disc and a planet of similar mass are disfavoured.

Finally, we found no CO or HCN line emission. These non-detections are
still consistent with icy planetesimals in HD~92945's debris disc with
volatile compositions similar to Solar System comets. Some of these
planetesimals could have been scattered in if the gap is of secondary
origin, cleared by a planet at 73~au and an additional planet is on an
orbit near the disc inner edge at 50~au.

\section*{Acknowledgements}

MB acknowledges support from the Deutsche Forschungsgemeinschaft (DFG)
through project Kr 2164/15-1. VF's postdoctoral fellowship is
supported by the Exoplanet Science Initiative at the Jet Propulsion
Laboratory, California Inst. of Technology, under a contract with the
National Aeronautics and Space Administration. GMK is supported by the
Royal Society as a Royal Society University Research Fellow. LM
acknowledges support from the Smithsonian Institution as a
Submillimeter Array (SMA) Fellow. This paper makes use of the
following ALMA data: ADS/JAO.ALMA\#2016.1.00104.S. ALMA is a
partnership of ESO (representing its member states), NSF (USA) and
NINS (Japan), together with NRC (Canada), MOST and ASIAA (Taiwan), and
KASI (Republic of Korea), in cooperation with the Republic of
Chile. The Joint ALMA Observatory is operated by ESO, AUI/NRAO and
NAOJ.





\bibliographystyle{mnras}
\bibliography{SM_pformation} 






\bsp	
\label{lastpage}
\end{document}